\documentclass[aps,pra,preprintnumbers,showpacs,tightenlines]{revtex4}
\usepackage{amssymb}
\usepackage{amsmath}
\usepackage{graphicx}
\usepackage{epsfig}
\usepackage{subfigure}
\usepackage{amsfonts}
\usepackage{CJK}

\begin{document}

\title{Deterministic transfer of multiqubit GHZ entangled states and quantum secret sharing between different cavities}

\author{Xiao-Ling He$^{1}$ and Chui-Ping Yang$^{2\star}$}

\address{$^1$School of Science, Zhejiang University of Science and Technology, Hangzhou, Zhejiang 310023, China}
\address{$^2$Department of Physics, Hangzhou Normal University, Hangzhou, Zhejiang 310036, China}
\address{$^\star$ yangcp@hznu.edu.cn}

\date{\today}

\begin{abstract}
We propose a way for transferring Greenberger-Horne-Zeilinger (GHZ)
entangled states from $n$ qubits in one cavity onto another $n$ qubits in the other cavity.
It is shown that $n$-qubit GHZ states $\alpha \left\vert 00...0\right\rangle +\beta
\left\vert 11...1\right\rangle $ with arbitrary degree of entanglement can
be transferred deterministically. Both of the GHZ state transfer
and the operation time are not dependent on the number of qubits, and there is no need
of measurement. This proposal is quite general and can be applied to
accomplish the same task for a wide range of physical qubits. Furthermore, note
that the $n$-qubit GHZ state $\alpha \left\vert 00...0\right\rangle
+\beta \left\vert 11...1\right\rangle $ is a quantum-secret-sharing code for
encoding a single-qubit arbitrary pure state $\alpha \left\vert
0\right\rangle +\beta \left\vert 1\right\rangle $. Thus, this work also
provides a way to transfer quantum secret sharing from $n$ qubits in one
cavity to another $n$ qubits in the other cavity.
\end{abstract}

\pacs{03.67.Bg, 42.50.Dv, 85.25.Cp, 76.30.Mi} \maketitle
\date{\today}

\section{Introduction}

Cavity-QED has been considered as one of the most powerful techniques for
quantum information processing (QIP). During the past years, a great amount
of work has been devoted to QIP with qubits coupled to (or placed in) a
single cavity. Attention has been recently shifting to large-scale QIP based
on cavity QED, which needs many qubits placed in different cavities. It is
noted that placing all of qubits in a single cavity can cause many
fundamental problems such as the increase in cavity decay rate and decrease
in qubit-cavity coupling strength. Hence, future cavity-based QIP may
require quantum networks consisting of multiple cavities, each hosting and
coupled to multiple qubits. In this type of quantum network, transfer of
quantum information will not only happen among qubits in the same cavity but
also occur between different cavities.

Among a variety of multiqubit entangled states, Greenberger-Horne-Zeilinger
(GHZ) states [1] are the archetype of multiqubit entangled states, which are
especially of interest and have drawn considerable attention. They can be
used to test nonlocality of quantum mechanics [1] and have applications in
quantum metrology [2] and high-precision spectroscopy [3-5]. Moreover, GHZ
states are useful in quantum teleportation [6,7], entanglement swapping [8],
quantum cryptographic [9], and error correction protocols [10,11]. Over the
past decade, based on cavity or circuit QED, a number of methods have been
proposed for creating GHZ states with a wide range of physical systems such
as atoms [12-14], quantum dots [15,16], superconducting (SC)\ qubits
[17-20], and photons [21]. Moreover, experiments have demonstrated
eight-photon GHZ states [22,23], fourteen-ion GHZ states [24],
three-SC-qubit GHZ states (based on circuit QED) [25], five-SC-qubit GHZ
states (via capacitance coupling) [26], and three-qubit GHZ states in NMR
[27].

Quantum state transfer (QST) plays an essential role in quantum
communication and is important in QIP. During the past decade, a great deal
of efforts has been devoted to one-qubit QST, i.e., transferring an
arbitrary unknown one-qubit state $\alpha \left\vert 0\right\rangle +\beta
\left\vert 1\right\rangle $ ($\left\vert \alpha \right\vert ^{2}+\left\vert
\beta \right\vert ^{2}=1$). Based on cavity/circuit QED, many theoretical
proposals have been presented for implementing one-qubit QST in various
physical systems [28-37], and one-qubit QST has been experimentally
demonstrated with superconducting qubits [38,39] and spatially-separated
atoms in a network [40]. Moreover, during the past years, much attention has
been paid to quantum entanglement transfer (QET). Many proposals for
implementing multi-qubit QET via quantum teleportation protocols have been
presented [41-45], and schemes for realizing QET based on cavity QED or
circuit QED have been also proposed [46-48]. Furthermore, QET has been
experimentally demonstrated in linear optics [49,50].

Motivated by the above, we here consider a physical system consisting of two
cavities each hosting $n$ qubits and coupled to a coupler qubit. In the
following, we will propose a way to transfer an $n$-qubit GHZ state $\alpha
\left\vert 00...0\right\rangle +\beta \left\vert 11...1\right\rangle $ (with
arbitrary unknown coefficients $\alpha $ and $\beta $) from $n$ qubits in
one cavity onto $n$ qubits in the other cavity. As shown below, this
proposal has the following features and advantages: (i) The proposal can be
used to implement the deterministic transfer of GHZ entangled states with
\textit{arbitrary degree of entanglement}, (ii) The GHZ state transfer does
not depend on the number of qubits, (iii) The operation time does not
increase as the number of qubits increases, (iv) No measurement is needed
during the operation, (v) The level $\left\vert f\right\rangle $ of only two
qubits is occupied during the operation, thus decoherence caused by energy
relaxation and dephasing from the qubits is much suppressed, and (vi) This
proposal is quite general and can be applied to a wide range of physical
qubits such as atoms, quantum dots, NV centers and various superconducting
qubits (e.g., phase, charge, flux, transmon, and Xmon qubits).

There are several additional motivations for this work, which are described
below:

First, the transfer of multiqubit entangled states is not only fundamental
in quantum mechanics but also important in QIP.

Second, multiqubit entangled states are essential resources for large-scale
QIP. When qubits in the two cavities belong to the same species,
transferring quantum entanglement is necessary in cavity-based large-scale
QIP, which is performed across different information processors each
consisting of a cavity and qubits in the cavity.

Third, when qubits in the two cavities are hybrid (i.e., different types),
qubits in one cavity can act as information process cells (i.e., the
operation qubits)\ while qubits in the other cavity play a role of quantum
memory elements (i.e., the memory qubits). When performing QIP, after a step
of information processing is completed, one may need to transfer quantum
states (either entangled or non-entangled)\ of the operation qubits\ (i.e.,
SC qubits, which are readily controlled and used for performing quantum
operations) to the memory qubits (i.e., NV centers [51] or atoms, which have
long decoherence time) for storage; and one needs to transfer the quantum
states from the memory qubits back to the operation qubits when a further
step of processing is needed. Note that hybrid quantum systems, composed of
different kinds of qubits (e.g., SC qubits and NV centers), have attracted
tremendous attentions recently and are considered as promising candidates
for QIP [52-55].

Last, according to [56], the $n$-qubit GHZ state $\alpha \left\vert
00...0\right\rangle +\beta \left\vert 11...1\right\rangle $ is a
quantum-secret-sharing code, which encodes a single-qubit arbitrary pure
state $\alpha \left\vert 0\right\rangle +\beta \left\vert 1\right\rangle $
via $n$ qubits. It is straightforward to see that after tracing over the
other qubits, the density operator for each qubit is an identity $I,$ i.e.,
the original quantum information carried by a single qubit is uniformly
distributed over $n$ qubits but each qubit does not carry any information.
For the detailed discussion, see [56]. Hence, the method presented here also
provides a way to transfer quantum secret sharing from $n $ qubits in one
cavity to another $n$ qubits in the other cavity.

After a deep literature search, we note that based on cavity/circuit QED,
how to transfer GHZ states between qubits distributed in different cavities
and how to transfer quantum secret sharing between different cavities have
not been reported.

This paper is outlined as follows. In Sec. II, we show a generic approach to
transfer $n$-qubit GHZ entangled states from one cavity to the other cavity.
In Sec. III, we give a brief discussion on the experimental issues. In Sec. IV,
we discuss the experimental feasibility of transferring a three-qubit GHZ
state in circuit QED. A concluding summary is given in Sec.~V.

\section{Transferring multi-qubit GHZ-state between two cavities}

\begin{figure}[tbp]
\begin{center}
\includegraphics[bb=71 533 517 655, width=8.5 cm, clip]{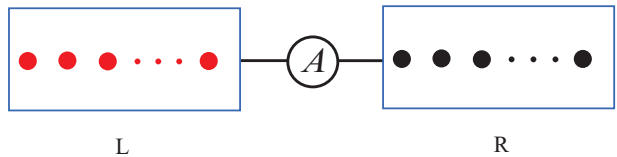} \vspace*{%
-0.08in}
\end{center}
\caption{(Color online) Diagram of two sets of qubits placed in two
different cavities connected to a coupler qubit. The circle $A$ in the
middle represents the coupler qubit (e.g., a superconducting qubit or a
quantum dot), which is capacitively or inductively coupled to each cavity.
Each red or dark dot represents a qubit. The red dots represent qubits
placed in the left cavity while the dark dots represent qubits placed in the
right cavity. Qubits in the same cavity are identical but qubits in
different cavities could be the same or non-identical (i.e., hybrid). In
addition, each square box indicates a cavity, which could be a
three-dimensional (3D) cavity or a one-dimensional (1D) cavity. The GHZ
states of qubits in one cavity can be transferred onto qubits in the other
cavity, as shown in the text.}
\label{fig:1}
\end{figure}

Consider two cavities $L$ and $R$ coupled to a coupler qubit $A$ and each
hosting $n$ qubits (Fig. 1). Without loss of generality, consider that
qubits in the same cavity are identical (e.g., SC qubits) but qubits in
different cavities are either the same or non-identity/hybrid (e.g., SC
qubits in cavity $L$ while NV centers in cavity $R$). The $n$ qubits in
cavity $L$ are labelled as $1,2,...,$ and $n$; while the $n$ qubits in
cavity $R$ are denoted as $1^{\prime },2^{\prime },...,n^{\prime }$. For
intra-cavity qubits, three levels $\left\vert g\right\rangle ,$ $\left\vert
e\right\rangle ,$ and $\left\vert f\right\rangle $ are employed, while for
the coupler qubit $A$ only two levels $\left\vert g\right\rangle _{A}$ and $%
\left\vert e\right\rangle _{A}$ are applied (Fig.~1). As shown below, the
GHZ state transfer employs the qubit-cavity resonant interaction and the
qubit-cavity dispersive interaction, which can be reached by adjusting the
level spacings of qubits [57-63].

The qubits and the coupler qubit are initially decoupled from their
respective cavities. Suppose that cavity $L$ ($R$) is initially in a vacuum
state $\left\vert 0\right\rangle _{L}$ $\left( \left\vert 0\right\rangle
_{R}\right) $, the coupler qubit is initially in the state $\left\vert
g\right\rangle _{A}$, the $n$ qubits ($1,2,...,n$) in cavity $L$ are
initially in a GHZ state
\begin{equation}
\left\vert GHZ\right\rangle _{12...n}=\alpha \left\vert g\right\rangle
_{1}\prod_{l=2}^{n}\left\vert +\right\rangle _{l}+\beta \left\vert
f\right\rangle _{1}\prod_{l=2}^{n}\left\vert -\right\rangle _{l}
\end{equation}
(with unknown coefficients $\alpha $ and $\beta $), and the $n$ qubits ($%
1^{\prime },2^{\prime },...,n^{\prime }$) in cavity $R$ are initially in the
state $\left\vert g\right\rangle _{1^{\prime }}\prod_{l^{\prime }=2^{\prime
}}^{n^{\prime }}\left\vert +\right\rangle _{l^{\prime }}.$ Here, $\left\vert
\pm \right\rangle =\left( \left\vert g\right\rangle \pm \left\vert
e\right\rangle \right) /\sqrt{2}$ are two orthogonal states. The initial
state of the whole system is thus given by

\begin{equation}
\left( \alpha \left\vert g\right\rangle _{1}\prod_{l=2}^{n}\left\vert
+\right\rangle _{l}+\beta \left\vert f\right\rangle
_{1}\prod_{l=2}^{n}\left\vert -\right\rangle _{l}\right) \otimes \left\vert
g\right\rangle _{A}\otimes \left\vert g\right\rangle _{1^{\prime
}}\prod_{l^{\prime }=2^{\prime }}^{n^{\prime }}\left\vert +\right\rangle
_{l^{\prime }}\otimes \left\vert 0\right\rangle _{L}\left\vert
0\right\rangle _{R}.
\end{equation}

In the following, the Hamiltonians are written in the interaction picture, $%
a^{+}$ ($b^{+}$) is the photon creation operator of cavity $L$ ($R$), and $%
\omega _{a\text{ }}$($\omega _{b}$) is the frequency of cavity $L$ ($R$).
The whole procedure for transferring the GHZ state of the $n$ qubits ($%
1,2,...,n$) in cavity $L$ onto the $n$ qubits ($1^{\prime},2^{%
\prime},...,n^{\prime}$) in cavity $R$ is listed below:

Step 1: Adjust the level spacings of qubit 1 to bring the $\left\vert
e\right\rangle \leftrightarrow \left\vert f\right\rangle $ transition on
resonance with cavity $L$ [Fig. 2(a)]. The Hamiltonian is given by $%
H_{1,1}=\hbar \left( \mu _{1}a^{+}\left\vert e\right\rangle _{1}\left\langle
f\right\vert +h.c.\right) ,$ where $\mu _{1}$ is the resonant coupling
strength between cavity $L$ and the $\left\vert e\right\rangle
\leftrightarrow \left\vert f\right\rangle $ transition of qubit 1. Under the
Hamiltonian $H_{1,1}$ and after an interaction time\ $t_{1,1}=\pi /\left(
2\mu _{1}\right) ,$ the state\ component $\left\vert f\right\rangle
_{1}\left\vert 0\right\rangle _{L}$ changes to $-i\left\vert e\right\rangle
_{1}\left\vert 1\right\rangle _{L}$ (for the details, see [64]). Now adjust
the level spacings of qubit $1$ to bring the $\left\vert g\right\rangle
\leftrightarrow \left\vert e\right\rangle $ transition on resonance with
cavity $L$ [Fig. 2(b)]. The Hamiltonian is $H_{1,2}=\hbar \left( \widetilde{%
\mu }_{1}a^{+}\left\vert g\right\rangle _{1}\left\langle e\right\vert
+h.c.\right) ,$ with $\widetilde{\mu }_{1}$ being the resonant coupling
strength between cavity $L$ and the $\left\vert g\right\rangle
\leftrightarrow \left\vert e\right\rangle $ transition of qubit 1. Under the
Hamiltonian $H_{1,2}$ and after an interaction time\ $t_{1,2}=\pi /\left( 2%
\sqrt{2}\widetilde{\mu }_{1}\right) ,$ the state\ component $\left\vert
e\right\rangle _{1}\left\vert 1\right\rangle _{L}$ changes to $-i\left\vert
g\right\rangle _{1}\left\vert 2\right\rangle _{L}$ [64]$.$

After this step of operation, we can obtain the transformation $\left\vert
f\right\rangle _{1}\left\vert 0\right\rangle _{L}\rightarrow -\left\vert
g\right\rangle _{1}\left\vert 2\right\rangle _{L}$ but the state component $%
\left\vert g\right\rangle _{1}\left\vert 0\right\rangle _{L}$ remains
unchanged because of $H_{1,1}\left\vert g\right\rangle _{1}\left\vert
0\right\rangle _{L}=H_{1,2}\left\vert g\right\rangle _{1}\left\vert
0\right\rangle _{L}=0.$ Thus, the initial state (2) of the whole system
becomes
\begin{equation}
\left\vert g\right\rangle _{1}\left( \alpha \prod_{l=2}^{n}\left\vert
+\right\rangle _{l}\left\vert 0\right\rangle _{L}-\beta
\prod_{l=2}^{n}\left\vert -\right\rangle _{l}\left\vert 2\right\rangle
_{L}\right) \otimes \left\vert g\right\rangle _{A}\otimes \left\vert
g\right\rangle _{1^{\prime }}\prod_{l^{\prime }=2^{\prime }}^{n^{\prime
}}\left\vert +\right\rangle _{l^{\prime }}\otimes \left\vert 0\right\rangle
_{R}.
\end{equation}

\begin{figure}[tbp]
\begin{center}
\includegraphics[bb=84 261 578 746, width=8.5 cm, clip]{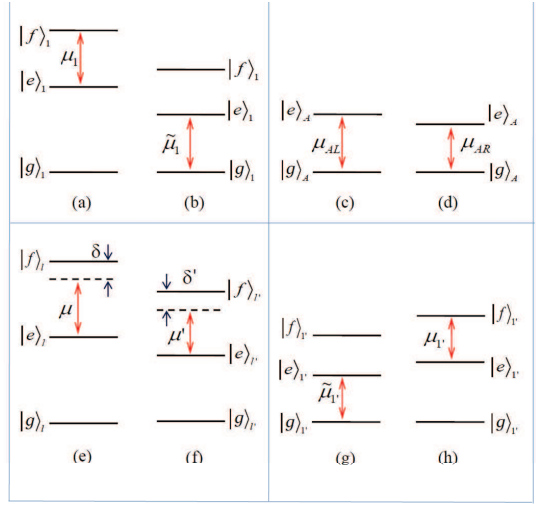} \vspace*{%
-0.08in}
\end{center}
\caption{(Color online) Illustration of qubit-cavity interaction. (a)
Resonant interaction between cavity $L$ with the $\left\vert e\right\rangle
\leftrightarrow \left\vert f\right\rangle $ transition of qubit 1. (b)
Resonant interaction between cavity $L$ with the $\left\vert g\right\rangle
\leftrightarrow \left\vert e\right\rangle $ transition of qubit 1. (c)
Resonant interaction between cavity $L$ with the $\left\vert g\right\rangle
\leftrightarrow \left\vert e\right\rangle $ transition of the coupler qubit $%
A$. (d) Resonant interaction between cavity $R$ with the $\left\vert
g\right\rangle \leftrightarrow \left\vert e\right\rangle $ transition of the
coupler qubit $A$. (e) Dispersive interaction between cavity $L$ and the $%
\left\vert e\right\rangle \leftrightarrow \left\vert f\right\rangle $
transition of qubits ($2,3,...,n$). In (e), the subscript $l=2,3,...,n$. (f)
Dispersive interaction between cavity $R$ and the $\left\vert e\right\rangle
\leftrightarrow \left\vert f\right\rangle $ transition of qubits ($2^{\prime
},3^{\prime},...,n^{\prime }$). In (f), the subscript $l^{\prime}=2^{\prime
},3^{\prime},...,n^{\prime }$. (g) Resonant interaction between cavity $R$
and the $\left\vert g\right\rangle \leftrightarrow \left\vert e\right\rangle
$ transition of qubit $1^{\prime }$. (h) Resonant interaction between cavity
$R $ and the $\left\vert e\right\rangle \leftrightarrow \left\vert
f\right\rangle $ transition of qubit $1^{\prime }$. When going from (a) to
(h), the level spacings of qubits and the coupler qubit need to be adjusted
to obtain the required qubit-cavity resonant or dispersive interaction. Note
that the level spacings of qubits and the coupler qubit can be readily
adjusted by varying the external control parameters or control fields (see,
[57-63]). In (a-h), each vertical blue-color arrow line indicates the mode
frequency of cavity $L$, while each vertical red-color arrow line represents
the mode frequency of cavity $R$.}
\label{fig:2}
\end{figure}

Step 2: Adjust the level spacings of qubit $1$ back to the previous
situation such that cavity $L$ is decoupled from this qubit. In the
meantime, bring the coupler qubit $A$ on resonance with cavity $L$ [Fig.
2(c)]. The Hamiltonian is $H_{2,1}=\hbar \left( \mu _{AL}a^{+}\left\vert
g\right\rangle _{A}\left\langle e\right\vert +h.c.\right) ,$ where $\mu
_{AL} $ is the resonant coupling strength between cavity $L$ and the coupler
qubit $A$. Under the Hamiltonian $H_{2,1}$ and after an interaction time\ $%
t_{2,1}=\pi /\left( 2\sqrt{2}\mu _{AL}\right) ,$ the state\ component $%
\left\vert g\right\rangle _{A}\left\vert 2\right\rangle _{L}$ changes to $%
-i\left\vert e\right\rangle _{A}\left\vert 1\right\rangle _{L}$ [64]. Now
bring the coupler qubit $A$ on resonance with cavity $R$ [Fig. 2(d)]. The
Hamiltonian is $H_{2,2}=\hbar \left( \mu _{AR}b^{+}\left\vert g\right\rangle
_{A}\left\langle e\right\vert +h.c.\right) ,$ where $\mu _{AR}$ is the
resonant coupling strength between cavity $R$ and the coupler qubit $A$.
Under the Hamiltonian $H_{2,2}$ and after an interaction time\ $t_{2,2}=\pi
/\left( 2\mu _{AR}\right) ,$ the state\ component $\left\vert e\right\rangle
_{A}\left\vert 1\right\rangle _{L}\left\vert 0\right\rangle _{R}$ changes to
$-i\left\vert g\right\rangle _{A}\left\vert 1\right\rangle _{L}\left\vert
1\right\rangle _{R}.$

After this step of operation, we can obtain the transformation $\left\vert
g\right\rangle _{A}\left\vert 2\right\rangle _{L}\left\vert 0\right\rangle
_{R}\rightarrow -\left\vert g\right\rangle _{A}\left\vert 1\right\rangle
_{L}\left\vert 1\right\rangle _{R}$ but the state component $\left\vert
g\right\rangle _{A}\left\vert 0\right\rangle _{L}\left\vert 0\right\rangle
_{R}$ remains unchanged due to $H_{2,1}\left\vert g\right\rangle
_{A}\left\vert 0\right\rangle _{L}\left\vert 0\right\rangle
_{R}=H_{2,2}\left\vert g\right\rangle _{A}\left\vert 0\right\rangle
_{L}\left\vert 0\right\rangle _{R}=0.$ Hence, the state (3) becomes
\begin{equation}
\left\vert g\right\rangle _{1}\left( \alpha \prod_{l=2}^{n}\left\vert
+\right\rangle _{l}\left\vert 0\right\rangle _{L}\left\vert 0\right\rangle
_{R}+\beta \prod_{l=2}^{n}\left\vert -\right\rangle _{l}\left\vert
1\right\rangle _{L}\left\vert 1\right\rangle _{R}\right) \otimes \left\vert
g\right\rangle _{A}\otimes \left\vert g\right\rangle _{1^{\prime
}}\prod_{l^{\prime }=2^{\prime }}^{n^{\prime }}\left\vert +\right\rangle
_{l^{\prime }}.
\end{equation}

Step 3: Bring the coupler qubit $A$ back to the original level configuration
such that the qubit $A$ is decoupled from the two cavities. Meanwhile,
adjust the level spacings of qubits ($2,3,...,n$) to have their $\left\vert
e\right\rangle \leftrightarrow \left\vert f\right\rangle $ transition
coupled to cavity $L$ [Fig. 2(e)], and adjust the level spacings of qubits ($%
2^{\prime },3^{\prime },...,n)$] to have their $\left\vert e\right\rangle
\leftrightarrow \left\vert f\right\rangle $ transition coupled to cavity $R$
[Fig. 2(f)]. The interaction Hamiltonian is given by

\begin{equation}
H=\sum_{l=2}^{n}\mu \left( e^{i\delta t}a\left\vert f\right\rangle
_{l}\left\langle e\right\vert +h.c.\right) +\sum_{l^{\prime }=2^{\prime
}}^{n^{\prime }}\mu ^{\prime }\left( e^{i\delta ^{\prime }t}b\left\vert
f\right\rangle _{l^{\prime }}\left\langle e\right\vert +h.c.\right) ,
\end{equation}%
where $\delta $$=\omega _{fe}-\omega _{a\text{ }},$ $\delta ^{\prime }$$%
=\omega _{fe}^{\prime }-\omega _{b\text{ }},$ and $\mu $ ($\mu ^{\prime }$)
is the non-resonant (dispersive) coupling strength between cavity $L$ ($R$)
and the $\left\vert e\right\rangle \leftrightarrow \left\vert f\right\rangle
$ transition of qubits ($2,3,...,n$) [qubits ($2^{\prime },3^{\prime },...,n)
$]. Here, $\omega _{fe}$ ($\omega _{fe}^{\prime }$) is the $\left\vert
e\right\rangle $ $\leftrightarrow $ $\left\vert f\right\rangle $ transition
frequency for qubits ($2,3,...,n$) [qubits ($2^{\prime },3^{\prime
},...,n^{\prime})$].

Under the large detuning condition $\delta \gg \mu $ and $\delta ^{\prime
}\gg \mu ^{\prime },$\ we can obtain the following effective Hamiltonian
[65,66]

\begin{eqnarray}
H &=&\lambda \sum\limits_{l=2}^{n}\left( \left\vert f\right\rangle
_{l}\left\langle f\right\vert aa^{+}-\left\vert e\right\rangle
_{l}\left\langle e\right\vert a^{+}a\right)  \notag \\
&&+\lambda ^{\prime }\sum\limits_{l^{\prime }=2^{\prime }}^{n^{\prime
}}\left( \left\vert f\right\rangle _{l^{\prime }}\left\langle f\right\vert
bb^{+}-\left\vert e\right\rangle _{l^{\prime }}\left\langle e\right\vert
b^{+}b\right)  \notag \\
&&+\lambda \sum_{l\neq k=2}^{n}\left\vert f\right\rangle _{l}\left\langle
e\right\vert \otimes \left\vert e\right\rangle _{k}\left\langle f\right\vert
\notag \\
&&+\lambda ^{\prime }\sum_{l^{\prime }\neq k^{\prime }=2}^{n}\left\vert
f\right\rangle _{l^{\prime }}\left\langle e\right\vert \otimes \left\vert
e\right\rangle _{k^{\prime }}\left\langle f\right\vert ,
\end{eqnarray}%
where $\lambda =\mu ^{2}/\delta $ and $\lambda ^{\prime }=\left( \mu
^{\prime }\right) ^{2}/\delta ^{\prime }$ are the effective coupling
strengths. The terms in lines 1 and 2 of Eq.~(6) describe the photon-number
dependent Stark shifts. The term in line 3 describes the \textquotedblleft
dipole\textquotedblright\ couplings between the $l$th qubit and the $k$th
qubit (in cavity $L$), and the term in the last line describes the
\textquotedblleft dipole\textquotedblright\ couplings between the $l^{\prime
}$th qubit and the $k^{\prime }$th qubit (in cavity $R$). Note that the
level $\left\vert f\right\rangle $ of each qubit is not involved in the
state (4). Thus, one can easily find that only the terms $-\lambda
\sum\limits_{l=2}^{n}\left\vert e\right\rangle _{l}\left\langle e\right\vert
a^{+}a$ and $-\lambda ^{\prime }\sum\limits_{l^{\prime }=2^{\prime
}}^{n^{\prime }}\left\vert e\right\rangle _{l^{\prime }}\left\langle
e\right\vert b^{+}b$ of Eq.~(6) have contribution to the time evolution of
the state (4), while all other terms in Eq.~(6) acting on the state (4)
result in zero. In other words, with respective to the state (4), the
Hamiltonian (6) reduces to
\begin{equation}
H=-\lambda \sum\limits_{l=2}^{n}\left\vert e\right\rangle _{l}\left\langle
e\right\vert a^{+}a-\lambda ^{\prime }\sum\limits_{l=2^{\prime }}^{n^{\prime
}}\left\vert e\right\rangle _{l^{\prime }}\left\langle e\right\vert b^{+}b.
\end{equation}%
Under the Hamiltonian (7), the state (4)\ evolves into

\begin{eqnarray}
&&\left\vert g\right\rangle _{1}\left( \alpha \prod_{l=2}^{n}\left\vert
+\right\rangle _{l}\prod_{l^{\prime }=2^{\prime }}^{n^{\prime }}\left\vert
+\right\rangle _{l^{\prime }}\left\vert 0\right\rangle _{L}\left\vert
0\right\rangle _{R}\right. +  \notag \\
&&\left. +\beta \prod_{l=2}^{n}\left( \left\vert g\right\rangle
_{l}-e^{i\lambda t}\left\vert e\right\rangle _{l}\right) \prod_{l^{\prime
}=2^{\prime }}^{n^{\prime }}\left( \left\vert g\right\rangle _{l^{\prime
}}+e^{i\lambda ^{\prime }t}\left\vert e\right\rangle _{l^{\prime }}\right)
\left\vert 1\right\rangle _{L}\left\vert 1\right\rangle _{R}\right)  \notag
\\
&&\otimes \left\vert g\right\rangle _{A}\otimes \left\vert g\right\rangle
_{1^{\prime }}.
\end{eqnarray}%
In the case of $t_{3}=\left( 2m+1\right) \pi /\lambda =\left( 2k+1\right)
\pi /\lambda ^{\prime }$ ($m$ and $k$ are zero or positive integers), we
have from Eq. (8)

\begin{equation}
\left\vert g\right\rangle _{1}\prod_{l=2}^{n}\left\vert +\right\rangle
_{l}\left( \alpha \prod_{l^{\prime }=2^{\prime }}^{n^{\prime }}\left\vert
+\right\rangle _{l^{\prime }}\left\vert 0\right\rangle _{L}\left\vert
0\right\rangle _{R}+\beta \prod_{l^{\prime }=2^{\prime }}^{n^{\prime
}}\left\vert -\right\rangle _{l^{\prime }}\left\vert 1\right\rangle
_{L}\left\vert 1\right\rangle _{R}\right) \otimes \left\vert g\right\rangle
_{A}\otimes \left\vert g\right\rangle _{1^{\prime }}.
\end{equation}

Step 4: Adjust the level structure of qubits ($2,3,...,n$) and qubits ($%
2^{\prime },3^{\prime },...,n^{\prime }$) back to the previous configuration
while bring the coupler qubit $A$ on resonance with cavity $L$ [Fig. 2(c)].
The Hamiltonian is given by $H_{2,1}$ above. Under the Hamiltonian $H_{2,1}$
and after an interaction time\ $t_{4,1}=\pi /\left( 2\mu _{AL}\right) ,$ the
state\ component $\left\vert g\right\rangle _{A}\left\vert 1\right\rangle
_{L}\left\vert 1\right\rangle _{R}$ changes to $-i\left\vert e\right\rangle
_{A}\left\vert 0\right\rangle _{L}\left\vert 1\right\rangle _{R}$. Then,
bring the coupler qubit $A$ on resonance with cavity $R$ [Fig. 2(d)]. The
Hamiltonian is given by $H_{2,2}$ above. Under the Hamiltonian $H_{2,2}$ and
after an interaction time\ $t_{4,4}=\pi /\left( 2\sqrt{2}\mu _{AR}\right) ,$
the state\ component $\left\vert e\right\rangle _{A}\left\vert
0\right\rangle _{L}\left\vert 1\right\rangle _{R}$ changes to $-i\left\vert
g\right\rangle _{A}\left\vert 0\right\rangle _{L}\left\vert 2\right\rangle
_{R}.$

After the operation of this step, we can get the transformation $\left\vert
g\right\rangle _{A}\left\vert 1\right\rangle _{L}\left\vert 1\right\rangle
_{R}\rightarrow -\left\vert g\right\rangle _{A}\left\vert 0\right\rangle
_{L}\left\vert 2\right\rangle _{R}$ but the state component $\left\vert
g\right\rangle _{A}\left\vert 0\right\rangle _{L}\left\vert 0\right\rangle
_{R}$ remains unchanged. Hence, the state (9) becomes

\begin{equation}
\left\vert g\right\rangle _{1}\prod_{l=2}^{n}\left\vert +\right\rangle
_{l}\otimes \left\vert g\right\rangle _{A}\otimes \left\vert 0\right\rangle
_{L}\otimes \left( \alpha \prod_{l^{\prime }=2^{\prime }}^{n^{\prime
}}\left\vert +\right\rangle _{l^{\prime }}\left\vert 0\right\rangle
_{R}-\beta \prod_{l^{\prime }=2^{\prime }}^{n^{\prime }}\left\vert
-\right\rangle _{l^{\prime }}\left\vert 2\right\rangle _{R}\right) \otimes
\left\vert g\right\rangle _{1^{\prime }}.
\end{equation}

Step 5: Bring the coupler qubit $A$ back to the original level configuration
such that it is decoupled from the two cavities. Meanwhile, adjust the level
spacings of qubit $1^{\prime }$ such that the $\left\vert g\right\rangle
\leftrightarrow \left\vert e\right\rangle $ transition of qubit $1^{\prime }$
is resonant with cavity $R$ [Fig. 2(g)]. The Hamiltonian is $H_{5,1}=\hbar
\left( \widetilde{\mu }_{1^{\prime }}b^{+}\left\vert g\right\rangle
_{1^{\prime }}\left\langle e\right\vert +h.c.\right) ,$ where $\widetilde{%
\mu }_{1^{\prime }}$ is the resonant coupling strength between cavity $R$
and the $\left\vert g\right\rangle \leftrightarrow \left\vert e\right\rangle
$ transition of qubit $1^{\prime }$. Under the Hamiltonian $H_{5,1}$ and
after an interaction time\ $t_{5,1}=\pi /\left( 2\sqrt{2}\widetilde{\mu }%
_{1^{\prime }}\right) ,$ the state\ component $\left\vert g\right\rangle
_{1^{\prime }}\left\vert 2\right\rangle _{R}$ changes to $-i\left\vert
e\right\rangle _{1^{\prime }}\left\vert 1\right\rangle _{R}$. Adjust the
level spacings of qubit $1^{\prime }$ so that the $\left\vert e\right\rangle
\leftrightarrow \left\vert f\right\rangle $ transition of qubit $1^{\prime }$
is resonant with cavity $R$ [Fig. 2(h)]. The Hamiltonian is given by $%
H_{5,2}=\hbar \left( \mu _{1^{\prime }}b^{+}\left\vert e\right\rangle
_{1^{\prime }}\left\langle f\right\vert +h.c.\right) ,$ with $\mu
_{1^{\prime }}$ being the resonant coupling strength between cavity $R$ and
the $\left\vert e\right\rangle \leftrightarrow \left\vert f\right\rangle $
transition of qubit $1^{\prime }$. Under this Hamiltonian and after an
interaction time\ $t_{5,2}=\pi /\left( 2\widetilde{\mu }_{1^{\prime
}}\right) ,$ the state\ component $\left\vert e\right\rangle _{1^{\prime
}}\left\vert 1\right\rangle _{R}$ changes to $-i\left\vert f\right\rangle
_{1^{\prime }}\left\vert 0\right\rangle _{R}.$

After performing this step of operation, we can get the transformation $%
\left\vert g\right\rangle _{1^{\prime }}\left\vert 2\right\rangle
_{R}\rightarrow -\left\vert f\right\rangle _{1^{\prime }}\left\vert
0\right\rangle _{R}$ but the state component $\left\vert g\right\rangle
_{1^{\prime }}\left\vert 0\right\rangle _{R}$ remains unchanged because of $%
H_{5,1}\left\vert g\right\rangle _{1^{\prime }}\left\vert 0\right\rangle
_{R}=H_{5,2}\left\vert g\right\rangle _{1^{\prime }}\left\vert
0\right\rangle _{R}=0.$ Thus, the state (10) of the whole system becomes
\begin{equation}
\left\vert g\right\rangle _{1}\prod_{l=2}^{n}\left\vert +\right\rangle
_{l}\otimes \left\vert g\right\rangle _{A}\otimes \left\vert 0\right\rangle
_{L}\left\vert 0\right\rangle _{R}\otimes \left( \alpha \left\vert
g\right\rangle _{1^{\prime }}\prod_{l^{\prime }=2^{\prime }}^{n^{\prime
}}\left\vert +\right\rangle _{l^{\prime }}+\beta \left\vert f\right\rangle
_{1^{\prime }}\prod_{l^{\prime }=2^{\prime }}^{n^{\prime }}\left\vert
-\right\rangle _{l^{\prime }}\right) .
\end{equation}%
After the operation, the level spacings of qubit $1^{\prime }$ needs to be
adjusted such that qubit $1^{\prime }$ is decoupled from cavity $R$.

Note that the last part of the product in Eq. (11) is the state of qubits ($%
1^{\prime },2^{\prime },...,n^{\prime }$), which is the same as the GHZ
state of qubits ($1,2,...,n$), described by Eq. (1). Thus, the original $n$%
-qubit GHZ state of qubits ($1,2,...,n$) in cavity $L$ has been transferred
onto qubits ($1^{\prime },2^{\prime },...,n^{\prime }$) in cavity $R$ after
the above operations. By applying classical pulse to qubit $1^{\prime },$
the states $\left\vert g\right\rangle _{1^{\prime }}$ and $\left\vert
f\right\rangle _{1^{\prime }}$ can be easily converted into the states $%
\left\vert +\right\rangle _{1^{\prime }}$ and $\left\vert -\right\rangle
_{1^{\prime }}$, respectively.

The irrelevant qubits in each step described above need to be decoupled from
their respective cavities. This requirement can be achieved by the
adjustment of the level spacings of the qubits. For example, (i) The level
spacings of superconducting qubits can be rapidly adjusted by varying
external control parameters (e.g. the magnetic flux applied to a
superconducting loop of phase, transmon, Xmon or flux qubits; see e.g.
[57-60]); (ii) The level spacings of NV centers can be readily adjusted by
changing the external magnetic field applied along the crystalline axis of
each NV center [61,62]; and (iii) The level spacings of atoms/quantum dots
can be adjusted by changing the voltage on the electrodes around each
atom/quantum dot [63].

Additional points may need to be addressed. First, because the same detuning
$\delta $ ($\delta ^{\prime }$) is set for qubits ($2,3,...,n$) [qubits ($%
2^{\prime },3^{\prime },...,n^{\prime }$)], the level spacings for qubits ($%
2,3,...,n$) [qubits ($2^{\prime },3^{\prime },...,n^{\prime }$)] can be
synchronously adjusted, e.g., via changing the common external control
parameters. Second, as shown above, the level $\left\vert f\right\rangle $
for qubits ($2,3,...,n$) and qubits ($2^{\prime },3^{\prime },...,n^{\prime
} $) is unpopulated, i.e., the level $\left\vert f\right\rangle $ is
occupied only for two qubits $1$ and $1^{\prime }$ ; thus decoherence from $%
2n-2$ qubits out of $2n$ qubits is greatly suppressed during the entire
operation. Third, the operation has nothing to do with $\alpha $ and $\beta ,
$ thus GHZ states with \textit{arbitrary degree of entanglement} can be
transferred by using this proposal. Last, the method is applicable to 1D, 2D
or 3D cavities or resonators as long as the conditions described above are
met.

Before ending this section, it should be pointed out that all
above-mentioned qubit-cavity resonant interactions involved during the GHZ
state transfer can be completed within a very short time, e.g., by
increasing the qubit-cavity resonant coupling strengths.

%\begin{center}
%\textbf{III. DISCUSSION}
%\end{center}

\section{Disscussion}

For the method to work, the following requirements need to be satisfied:

(i) The condition $\left( 2m+1\right) \pi /\lambda =\left( 2k+1\right) \pi
/\lambda ^{\prime }$ needs to be met. Because of $\lambda =\mu ^{2}/\delta $
and $\lambda ^{\prime }=\left( \mu ^{\prime }\right) ^{2}/\delta ^{\prime },$
this condition can be readily reached with an appropriate choice of $\delta $
(or $\delta ^{\prime }$) via adjusting the level spacings of qubits ($%
2,3,...,n$) [or qubits ($2^{\prime },3^{\prime },...,n^{\prime }$)]. For the
case when qubits in the two cavities belong to the same species and the two
cavities are identical, one would have\ $\lambda =\lambda ^{\prime }$ (i.e.,
$\delta =\delta ^{\prime }$ and $\mu =\mu ^{\prime }$) and thus could choose
$m=k=0$ to have $\tau _{5}=\pi /\lambda =\pi /\lambda ^{\prime }$, i.e., the
shortest operation time for step 3.

(ii) During step 3, the occupation probability $p$ of the level $\left\vert
f\right\rangle $ for each of qubits ($2,3,...,n$) and the occupation
probability $p^{\prime }$ of the level $\left\vert f\right\rangle $ for each
of qubits ($2^{\prime },3^{\prime },...,n^{\prime }$) are given by [67,68]
\begin{equation}
p\simeq \frac{4\mu ^{2}}{4\mu ^{2}+\delta ^{2}},\text{ }p^{\prime }\simeq
\frac{4\left( \mu ^{\prime }\right) ^{2}}{4\left( \mu ^{\prime }\right)
^{2}+(\delta ^{\prime })^{2}}.
\end{equation}%
The occupation probabilities $p$ and $p^{\prime }$ need to be negligibly
small in order to reduce the operation error. With the choice of $\delta
=10\mu $ and $\delta ^{\prime }=10\mu ^{\prime },$ one has $p,$ $p^{\prime
}\sim 0.04$, which can be further reduced by increasing the ratio of $\delta
/\mu $ and $\delta ^{\prime }/\mu ^{\prime }.$

(iii) The total operation time is
\begin{equation}
\tau =\tau _{r}+\tau _{o}+\tau _{a},
\end{equation}%
with
\begin{eqnarray}
\tau _{r} &=&\frac{\pi }{2}\left( \mu _{1}^{-1}+\mu _{1^{\prime }}^{-1}+\mu
_{AL}^{-1}+\mu _{AR}^{-1}\right) +\frac{\pi }{2\sqrt{2}}\left( \widetilde{%
\mu }_{1}^{-1}+\widetilde{\mu }_{1^{\prime }}^{-1}+\mu _{AL}^{-1}+\mu
_{AR}^{-1}\right) , \\
\tau _{o} &=&\left( 2m+1\right) \pi /\lambda =\left( 2k+1\right) \pi
/\lambda ^{\prime }, \\
\tau _{a} &=&6\tau _{A}+3\tau _{1}+3\tau _{1^{\prime }}+2\tau _{q}+2\tau
_{q^{\prime }}.
\end{eqnarray}%
Here, $\tau _{r}$ is a total of resonance operation time for steps 1, 2, 4,
and 5; $\tau _{o}$ is the off-resonance operation time for step 3; and $\tau
_{a}$ is a total of time required for adjusting the level spacings of the
qubits and the coupler qubit. In addition, $\tau _{1},$ $\tau _{1^{\prime }},
$ and $\tau _{A}$ are the typical times needed for adjusting the level
spacings of qubit 1, qubit $1^{\prime },$ and the coupler qubit $A$,
respectively; $\tau _{q}$ ($\tau _{q^{\prime }}$) is the typical time
required for adjusting the level spacings of qubits ($2,3,...,n$) [qubits ($%
1^{\prime },2^{\prime },...,n^{\prime }$)].

From Eqs.~(13-16), one can see that the operation time $\tau $ is
independent of the number of qubits. To reduce decoherence, the operation
time $\tau $ should be much smaller than the energy relaxation time and the
dephasing time of qubits. In addition, $\tau $ should be much smaller than
the lifetime of the cavity mode, which is given by $\kappa
_{j}^{-1}=Q_{j}/\omega _{j}$ ($j=a,b$). Here, $Q_{a}$ ($Q_{b}$) is the
quality factor of cavity $L$ ($R$). In principle, these requirements can be
satisfied. The $\tau _{r}$ can be reduced by increasing the resonant
coupling strengths $\mu _{1},\widetilde{\mu }_{1},\mu _{1^{\prime }},%
\widetilde{\mu }_{1^{\prime }},\mu _{AL},$ and $\mu _{AR}.$ The $\tau _{a}$
can be reduced by rapidly adjusting the level spacings of the qubits and the
coupler qubit (e.g., $1-3$ ns is the typical time for adjusting the level
spacings of superconducting qubits in experiment [69,70]). And, $\kappa
^{-1} $ can be increased by employing high-$Q$ cavities.

\section{Possible experimental implementation}

As an example, let us give a discussion of the experimental possibility of
transferring a three-qubit GHZ state from three identical superconducting
transmon qubits in one cavity to another three identical superconducting
transmon qubits in the other cavity (Fig.~3). Each cavity considered here is
a one-dimensional transmission line resonator (TLR), and the two cavities
are coupled to a superconducting transmon qubit (Fig. 3).

\begin{figure}[tbp]
\begin{center}
\includegraphics[bb=141 479 475 579, width=9.5 cm, clip]{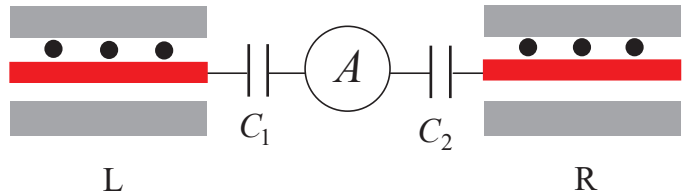} \vspace*{%
-0.08in}
\end{center}
\caption{(Color online) Setup for two cavities $L$ and $R$ coupled by a
superconducting transmon qubit $A$. Each cavity here is a one-dimensional
coplanar waveguide transmission line resonator. The circle $A$ represents a
superconducting transmon qubit (i.e., an artificial atom), which is
capacitively coupled to cavity $L$ ($R$) via a capacitance $C_1$ ($C_2$).
Each dark dot indicate an intra-cavity superconducting transmon qubit.}
\label{fig:3}
\end{figure}

Assume $\widetilde{\mu }_{1}\sim \widetilde{\mu }_{1^{\prime }}\sim \mu
_{AL}\sim \mu _{AR}=g$ and $g=2\pi \times 50$ MHz. The coupling strengths
with the values chosen here are readily available in experiments because a
coupling strength $g/2\pi \sim 360$ MHz has been reported for a transmon
qubit coupled to a TLR [71,72]. For a transmon qubit, one has $\mu _{1}\sim
\sqrt{2}\widetilde{\mu }_{1}$ and $\mu _{1^{\prime }}\sim \sqrt{2}\widetilde{%
\mu }_{1^{\prime }}$ [73], and thus $\mu _{1}\sim \mu _{1^{\prime }}\sim
2\pi \times 71$ MHz. For the coupling strengths chosen here, we have $\tau
_{r}\sim 31.2$ ns. For $\tau _{A}\sim \tau _{1}\sim \tau _{1^{\prime }}\sim
\tau _{q}\sim \tau _{q^{\prime }}=3$ ns, we have $\tau _{a}\sim 48$ ns. On
the other hand, as a rough estimate, assume $\mu \sim \mu _{1}\sim 2\pi
\times 71$ MHz, $\mu ^{\prime }\sim \mu _{1}^{\prime }\sim 2\pi \times 71$
MHz, $\delta \sim 10\mu ,$ and $\delta ^{\prime }\sim 10\mu ^{\prime }$. As
a result, we have $\tau _{o}=\pi \delta /\mu ^{2}=\pi \delta ^{\prime }/\mu
^{\prime 2}\sim 71.4$ ns. Hence, the total operation time $\tau =\tau
_{r}+\tau _{o}+\tau _{a}$ would be $\sim 0.15$ $\mu $s, which is much
shorter than the experimentally-reported energy relaxation time $T_{1}$ and
dephasing time $T_{2}$ of the level $\left\vert e\right\rangle $ and the
energy relaxation time $T_{1}^{\prime }$ and dephasing time $T_{2}^{\prime }$
of the level $\left\vert f\right\rangle $ of the transmon qubit. This is
because: (i) For a transmon qubit, $T_{1}^{\prime }\sim T_{1}/2$ and $%
T_{2}^{\prime }\sim T_{2}$ [69]; and (ii) $T_{1}$ and $T_{2}$ can be made to
be on the order of $20-60$ $\mu $s for state-of-the-art superconducting
transmon devices at the present time [74-76]. For a transmon qubit, the
typical transition frequency between two neighbor levels $\left\vert
e\right\rangle $ and $\left\vert f\right\rangle $ is $1-20$ GHz. As an
example, choose $\omega _{fe}/2\pi =\omega _{fe}^{\prime }/2\pi =10.0$ GHz.
For the values of $\mu $ and $\mu ^{\prime }$ given above, we have $\delta
/2\pi \sim \delta ^{\prime }/2\pi \sim 707$ MHz, and thus $\omega _{a}/2\pi
=\omega _{b}/2\pi \sim 9.293$ GHz. In addition, consider $Q_{a}=Q_{b}\sim
3\times 10^{5},$ and thus we have $\kappa _{a}^{-1}=\kappa _{b}^{-1}\sim 5.1$
$\mu $s, which is much longer than the operation time $\tau \sim 0.15$ $\mu $%
s given above. The required cavity quality factors here are achievable in
experiment because TLRs with a (loaded) quality factor $Q\sim 10^{6}$ have
been experimentally demonstrated [77,78]. The result presented here shows
that transferring three-qubit GHZ states between two TLRs is possible within
present-day circuit QED. We remark that further investigation is needed for
each particular experimental setup. However, this requires a rather lengthy
and complex analysis, which is beyond the scope of this theoretical work.

%\begin{center}
%\textbf{IV. CONCLUSION}
%\end{center}

\section{Conclusion}

We have shown that $n$-qubit GHZ states (with an arbitrary degree of
entanglement) can be transferred from $n$ qubits in one cavity to another $n$
qubits in the other cavity. This approach has several distinguishing
advantages mentioned in the introduction. We have given a discussion of the
experimental issues and provided an analysis on the experimental feasibility
of transferring a three-qubit GHZ states between two cavities within circuit
QED. The method presented here is quite general and can be applied to a wide
range of physical systems. This work is of interest because it is the first
to show that multi-qubit GHZ states or quantum secret sharing can be
transferred from one cavity to the other cavity, which is fundamental in
quantum mechanics and of importance in large-scale QIP and quantum
communication.

\section{Acknowledgments}

This work was supported in part by the National Natural Science Foundation
of China under Grant Nos. 11074062 and 11374083, the Zhejiang Natural
Science Foundation under Grant No. LZ13A040002, and the funds from Hangzhou
Normal University under Grant Nos. HSQK0081 and PD13002004. This work was
also supported by the funds of Hangzhou City for supporting the
Hangzhou-City Quantum Information and Quantum Optics Innovation Research
Team.

\end{document}